\documentclass[letterpaper, 10 pt, conference]{ieeeconf}  % Comment this line out if you need a4paper

\IEEEoverridecommandlockouts                              % This command is only needed if 
\usepackage{cite}
\usepackage{mathtools}
\usepackage{amsmath,amssymb,amsfonts}
\usepackage[english]{babel}
\newtheorem{definition}{Definition}[section]

\newtheorem{lemma}[definition]{Lemma}

\newtheorem{thm}{Theorem}[section]

\newtheorem{remark}{Remark}

\usepackage{algorithmic,balance}
\usepackage{graphicx}
\usepackage{textcomp}
\usepackage{xcolor}

\title{\LARGE \bf
Learning Neural Networks under Input-Output Specifications
\thanks{$^{1}$The Bradley Department of Electrical and Computer Engineering, Virginia Tech, Blacksburg, VA, USA. 
        {\tt\small Emails: \{zabdeen, kekatos, jinming\}@vt.edu}}%
\thanks{$^{2}$Department of Mechanical Engineering, University of California, Berkeley, CA, USA.
        {\tt\small Email: he\_yin@berkeley.edu}}%
        \thanks{This work was supported by the U.S. National Science Foundation under grant 2034137. Partial support of this research was provided by a funding by C3.ai and Microsoft, and the Woodrow W. Everett, Jr. SCEEE Development Fund in cooperation with the Southeastern Association of Electrical Engineering Department Heads.}}%

\author{Zain ul Abdeen$^{1}$, He Yin$^2$, Vassilis Kekatos$^1$, Ming Jin$^{1}$}

\begin{document}

\maketitle
\thispagestyle{empty}
\pagestyle{empty}

%%%%%%%%%%%%%%%%%%%%%%%%%%%%%%%%%%%%%%%%%%%%%%%%%%%%%%%%%%%%%%%%%%%%%%%%%%%%%%%%
\begin{abstract}
In this paper, we examine an important problem of learning neural networks that certifiably meet certain specifications on input-output behaviors.  Our strategy is to find an inner approximation of the set of admissible policy parameters, which is convex in a transformed space. To this end, we address the key technical challenge of convexifying the verification condition for neural networks, which is derived by abstracting the nonlinear specifications and activation functions with quadratic constraints. In particular, we propose a reparametrization scheme of the original neural network based on loop transformation, which leads to a convex condition that can be enforced during learning. This theoretical construction is validated in an experiment that specifies reachable sets for different regions of inputs.  
\end{abstract}

%%%%%%%%%%%%%%%%%%%%%%%%%%%%%%%%%%%%%%%%%%%%%%%%%%%%%%%%%%%%%%%%%%%%%%%%%%%%%%%%
	\section{Introduction}
The advances in deep learning (DL) have impacted many areas, such as computer vision and natural language processing \cite{goodfellow2016deep,sutskever2014sequence,krizhevsky2012imagenet}. However, the use of DL for safety-critical tasks in the real world is challenged by its opaqueness, fragility, and vulnerability~\cite{stoica}. For example, an imperceptible but carefully engineered perturbation in the input can easily mislead DL systems~\cite{szegedy2013intriguing}. Notably, DL models are rarely used in a standalone manner but as part of a larger pipeline. Thus, specifications on the model decisions are required to capture the true constraints on their physical and social ramifications. These specifications include but are not limited to safety~\cite{A1}, stability \cite{jin2020stability,yin2021imitation}, privacy~\cite{A3}, fairness~\cite{A4}, and interpretability~\cite{A5}. Up till now, verification of NN has been primarily focused on adversarial robustness, and can be divided into exact and inexact approaches. Exact approaches calculate the NN output set without any approximation, whereas inexact methods seek to approximate the output set for computational tractability~\cite{silva2020opportunities}. Moreover, deriving guarantees for nonlinear, large-scale, complex policies such as an NN is a significant technical challenge, and there have been increasing efforts towards this direction~\cite{gehr2018ai2,wong2018scaling,dvijotham2018training}.
	
The method that we propose for learning NN under specifications belongs to the large family of convex relaxation techniques. We demonstrate that for a specific class of specification problems, learning can be accomplished by solving a semidefinite program (SDP). In particular, we note that the integral quadratic constraint (IQC) framework~\cite{C1} has been applied in post-hoc verification of robustness for an already trained NN~\cite{fazlyab2020safety}. We address the main challenge of existing methods, that is the nonconvexity of the learning condition with respect to the policy parameters, to develop a computationally efficient procedure.
	
	\emph{Contribution:}
	We propose a framework to learn a NN that satisfies specifications on input-output behavior. We overcome a major technical hurdle by deriving a convex condition that can be imposed during the learning process. The key to our method is to compute a convex inner approximation to the nonconvex set of admissible policy parameters. To this end, we characterize the behaviors of the nonlinear activation and input-output specifications by quadratic constraints. For convexification, we design a new reparameterization scheme based on loop transformation \cite[Chap. 4]{sastry2013nonlinear} and $\mathcal{S}$-lemma~\cite{beta}, \cite{boyd1994linear}. The one-to-one correspondence between the transformed parameters and the original parameters are guaranteed for a two-layer NN. Hence, we can efficiently learn in the reparameterized space and recover the original parameters, leading to a NN that certifiably satisfies the desired properties.

	\emph{Related work:} IQC-based analysis of NN has been explored in \cite{fazlyab2020safety} under the verification setting. However, the corresponding condition is nonconvex in the policy parameters, preventing its application in a learning setting. As neural networks become popular in control tasks, safety and robustness of NN controlled systems have been examined in~\cite{yin2020recurrent,chow2018lyapunov,zhang2016learning,revay2020convex}. It is also possible to concatenate an optimization layer to a DNN to satisfy hard output constraints, with an additional computational cost to solve an optimization problem every time an output is produced~\cite{DC3}. An approach for NN verification against convex-relaxable specifications is proposed in \cite{qin2019verification}, which shares the line of thinking with the present work to move towards general specifications (beyond adversarial robustness). In \cite{pauli2021training}, a framework is designed based on SDP to measure the robustness of an input-output map defined by NN. These articles \cite{ref1,ref2} shed lights upon the convexification techniques for training NN. The present work is inspired by \cite{yin2021imitation}, in which the authors proposed a method to synthesize a NN controller with stability and safety guarantees through imitation learning. A recent extension to policy gradient for reinforcement learning is presented in \cite{yin2020recurrent}. 

The rest of the paper is organized as follows. Section~II describes the problem setup. Section~III briefly reviews results for verification of a fixed NN. The main method  to obtain a convex learning condition is presented in Section~IV. Section~V validates the approach in a reachability setup for different regions of inputs. Section~VI concludes the paper with some future directions.

\emph{Notation:} We denote $\mathbb{S}^{n}$, $\mathbb{S}^{n}_{+}$, $\mathbb{S}^{n}_{++}$ as the sets of $n \times n$ symmetric, positive semi-definite and positive definite matrices, respectively. For any matrix 
$A \in \mathbb{S}^{n}$ , the inequality  $A \succeq 0 $ and $A \succ 0$ indicates positive semi-definiteness and positive definiteness, respectively.

	%%%%%%%%%%%%%%%%%%%%%%%%%%%%%%%%%%%%%%%%%%%%%%%%%%%%%%%%%%%
	\section{Problem Formulation}
	\subsection{Problem statement}
	We consider specifications on outputs in relation to inputs that vary across instances. Formally, we define a multi-layer feed-forward neural network (NN) mapping $\Psi(\cdot;\theta):\mathcal{X}\to \mathcal{Y}$ parameterized by a weight vector $\theta \in \mathbb{R}^{n_\theta}$. Sets $\mathcal{X}\subseteq \mathbb{R}^{n_x}$ and $\mathcal{Y}\subseteq \mathbb{R}^{n_y}$ are respectively the sets wherein NN inputs and outputs can lie. We also define an $m$-way specification $F:\mathcal{X}\times \mathcal{Y}\to \mathbb{R}^{m}$, and its associated specification  set $\mathcal{S}(x)\coloneqq \{y\in\mathbb{R}^{n_y}: F(x,y)\leq 0\}$. Our aim is to find a  parameter $\theta$ such that 
	\begin{equation}\label{eq1}
	\Psi(x; \theta)\in \mathcal{S}(x),~~\forall x\in \mathcal{X}.
	\end{equation}
	To simplify notation, we henceforth omit the dependence of $\Psi$ on $\theta$. Note that the admissible set $\Theta \coloneqq \{\theta \in \mathbb{R}^{n_\theta}: \eqref{eq1} \  \text{is satisfied}\}$ is nonconvex in general due to the nonlinearity of $\Psi$ and specifications $F$. The above formulation can incorporate a family of problems in machine learning and control, such as fairness \cite{A4}, adversarial robustness~\cite{fazlyab2020safety,qin2019verification}; and reachability analysis \cite{fazlyab2020safety,hu2020reach}. Granted that searching within the nonconvex admissible set $\Theta$ is intractable, our strategy is to compute a convex inner approximation. To this end, we propose a semidefinite convexification approach to specify the convex set.
	
	\subsection{Isolating NN nonlinearities}
	The input-output mapping of a feed-forward NN with $l$ layers can be described by the recursive equations:
	\begin{equation}\label{nn1}
	\centering
	\begin{split}
	x^0&=x\\ x^{k+1}&=\phi^k(W^kx^k+b^k),~~~~~k=0,1,\dots,l-1\\ \Psi(x)&=W^lx^l+b^l
	\end{split}
	\end{equation}
	where $x\in \mathcal{X}$ is the NN input; $W^k \in \mathbb{R}^{n_{k+1}\times n_k}$ and $b^k\in \mathbb{R}^{n_{k+1}}$ are the weight matrix and bias vector of the $(k+1)$-th layer, respectively; and  $n=\sum_{k=1}^{l}n_k$  neurons. The mapping $\phi^k$ applies a nonlinear scalar activation function $\psi:\mathbb{R}\to \mathbb{R}$ on each one of the entries of its vector argument $W^kx^k+b^k$. The mapping can be defined as:
	\begin{eqnarray}
	\phi^k(x)=[\psi(x_1)~\psi(x_2)~\dots~ \psi(x_{n_k})]^{\top}.
	\end{eqnarray}
	Common choices for the scalar activation function include the hyperbolic tangent $\tanh$, the sigmoid, and the rectified linear unit (ReLU). The NN output $\Psi(x)$ is application-dependent. For example, in classification, $\Psi(x)$ is the logit input to a softmax function; in feedback control, $\Psi(x)$ is the control to the plant at state $x$. 
	
	To facilitate subsequent derivations, let us isolate the nonlinear and linear components of a NN as in \cite{fazlyab2020safety,yin2021imitation}. Let $v^k=W^k x^k +b^k$ denote the input to the activation function at layer $k+1$. Then, the NN defined in \eqref{nn1} can be rewritten as
	\begin{align}\label{A}
	\begin{bmatrix} v_\phi\\ \Psi(x)  \end{bmatrix}
	&=
	N\begin{bmatrix} x \\ x_\phi  \end{bmatrix}+ \begin{bmatrix} b_\phi \\ b^l\  \end{bmatrix}\\
	{x_\phi}&=\phi (v_\phi)
	\end{align}
where
$$\tiny{v_\phi = \begin{bmatrix} v^0 \\ \vdots \\ v^{l-1} \end{bmatrix}, x_\phi = \begin{bmatrix} x^1 \\ \vdots \\ x^{l} \end{bmatrix}, b_\phi = \begin{bmatrix} b^0 \\ \vdots \\ b^{l-1} \end{bmatrix}, \phi(v_\phi) = \begin{bmatrix} \phi^0(v^0) \\ \vdots \\ \phi^{l-1}(v^{l-1})\end{bmatrix}}.$$\color{black}
Matrix $N$ depends on the NN weights and can be partitioned as follows:
\begin{align*}
 N&= \small{\left[ 
\begin{array}{c | c}
\begin{array}{c}
     W^{0}  \\
     0\\
     \vdots
\end{array}&
\begin{array}{c c c c c} 
		 0 & \dots&0&0&0\\ 
		 W^{1} & \dots& 0&0&0\\ 
		\vdots &\dots& \vdots & W^{l-1}&0 
	\end{array} \\	 
	\hline 
	\begin{array}{c}
		0
	\end{array}&
	\begin{array}{ccccc}
	~~~~0&\dots&0&~~~0&~~~~W^{l}
	\end{array}
\end{array} 
\right]} \\
& 	=\begin{bmatrix}N_{vx} & N_{vx^1} \\ N_{\Psi x} & N_{\Psi x^{1}}  \end{bmatrix}. 
\end{align*}

\section{Specification Analysis for a Fixed NN}
We now briefly review the analysis conducted in \cite{fazlyab2020safety} based on the framework of quadratic constraints. For this section, the NN parameters are assumed already learned and fixed.

	%%%%%%%%%%%%%%%%%%%%%%%%%
	\subsection{Input set}
	We focus on the type of input sets that can be characterized as follows~\cite{fazlyab2020safety,hu2020reach}.
	
	\begin{definition}[Quadratic Constraints]
		Let $\mathcal{X} $ be a nonempty set and $\mathcal{P} \subset \mathbb{S}^{n_{x}+1}$  be the set of all symmetric (but possibly indefinite matrices) $P$ such that the following quadratic constraint (QC) holds for all $x\in \mathcal{X}$:
		\begin{gather}\label{defQC}
		\begin{bmatrix} x  \\ 1 \end{bmatrix}^\top
		P
		\begin{bmatrix} x \\ 1 \end{bmatrix}
		\geq 0.
		\end{gather}
		Then, we say that $\mathcal{X}$ satisfies the QC defined by $\mathcal{P}$. 
	\end{definition}
	
	\begin{remark}
		By definition, set $\mathcal{P}$ is a convex cone as for any $P_1,P_2\in \mathcal{P}$, we have $c_1P_1+c_2P_2\in \mathcal{P}$ for any $c_1,c_2\in \mathbb{R}^{+}$.
	\end{remark}
	
	Thus, we can over-approximate $\mathcal{X}$ with the intersection of possibly infinite number of sets:
	\begin{gather}
	\mathcal{X} \subseteq \bigcap_{P\in \mathcal{P}} \left\{ x\in \mathbb{R}^{n_{x}} :
	\begin{bmatrix} x  \\ 1 \end{bmatrix}^\top
	P
	\begin{bmatrix} x \\ 1 \end{bmatrix}
	\geq 0\right\}
	\end{gather}
	
	Our focus here is on ellipsoidal input sets  $\mathcal{X}=\left\lbrace x\in \mathbb{R}^{n_{x}} : \left\| Ax+b\right\| _{2}\leq 1 \right\rbrace  $ with $A\in \mathbb{S}^{n_{x}}$ and $b\in \mathbb{R}^{n_{x}}$, which satisfies QC with
		\begin{gather}
		\mathcal{P}_{\mathcal{X}}=\left\lbrace \lambda\begin{bmatrix}
		-A^TA& A^Tb\\-b^TA & 1-b^Tb
		\end{bmatrix}, ~~\lambda\geq 0\right\rbrace.
		\end{gather}  
Other types of sets such as polytopes, hyper-rectangles, and zonotopes can be also incorporated as in~\cite{fazlyab2020safety,hu2020reach}.

% 	\begin{itemize}
% 		\item Ellipsoids:  $\mathcal{X}=\left\lbrace x\in \mathbb{R}^{n_{x}} : \left\| Ax+b\right\| _{2}\leq 1 \right\rbrace  $, is an ellipsoid with $A\in \mathbb{S}^{n_{x}}$ and $b\in \mathbb{R}^{n_{x}}$, which satisfies QC with
% 		\begin{gather}
% 		\mathcal{P}_{\mathcal{X}}=\left\lbrace \lambda\begin{bmatrix}
% 		-A^TA& A^Tb\\-b^TA & 1-b^Tb
% 		\end{bmatrix}, ~~\lambda\geq 0\right\rbrace.
% 		\end{gather} 
% 	\end{itemize}

\subsection{Specification set}
The desirable properties to be verified are formulated as a specification set $\mathcal{S}(x)$ in the output space of the NN. However checking the condition $\Psi(x)\in \mathcal{S}(x)$ for all $x\in \mathcal{X}$ is a challenging task as it requires an exact computation of the non-convex set of outputs. Instead, our goal is to find a non-conservative over-approximation to the output set and verify the safety properties with respect to the new set. We assume the safety set is represented by the intersection of finitely many quadratic inequalities as
	\begin{gather}
	\mathcal{S}(x)=  \bigcap_{i=1}^{m} \left\lbrace y\in \mathbb{R}^{n_{y}} :
	\begin{bmatrix} x \\ y \\ 1 \end{bmatrix}^\top
	S_{i}
	\begin{bmatrix} x \\ y \\1 \end{bmatrix}
	\leq 0\right\rbrace
	\end{gather}
	where matrices $S_{i}\in \mathbb{S}^{n_{x}+n_{y}+1}$ are given. For instance, if the output set is specified by an ellipsoid as $\mathcal{S}(x) = \left\lbrace y \in \mathbb{R}^{n_{y}} : \left\|Cy +d\right\|_{\textcolor{black}{2}}\leq 1 ,~ y = \Psi(x) \right\rbrace$, \textcolor{black}{with $C\in \mathbb{S}^{n_{y}}$ and $d\in \mathbb{R}^{n_{y}}$,} then we can choose 
	\begin{gather}
	S=\begin{bmatrix}
	0 & 0& 0\\0 &C^{\top}C & C^{\top}d\\0& d^{\top}C& d^{\top}d-1
	\end{bmatrix}.
	\end{gather}

	\subsection{Abstraction of activation functions}
	One of the challenges in the analysis of NN is the composition of nonlinear  activation functions. By exploiting the common patterns of activation functions, a viable approach is to employ sector bounds~\cite{yin2021imitation,fazlyab2020safety}. We begin with a formal definition.
	
	\begin{definition}[QC for functions] Let $\phi:\mathbb{R}^{n}\to\mathbb{R}^{n}$ and suppose $\mathcal{Q}\subset \mathbb{S}^{2n+1}$ is the set of all symmetric indefinite matrices $Q$ such that the  following inequality holds for all $x\in\mathbb{R}^{n}$: 
	\begin{gather}
	\begin{bmatrix} x \\ \phi(x) \\ 1 \end{bmatrix}^\top
	Q
	\begin{bmatrix} x \\ \phi(x) \\1 \end{bmatrix}
	\leq 0.
	\end{gather}
	Then, we say that $\phi$ satisfies the QC defined by $\mathcal{Q}$. 
	\end{definition}
	
	The derivation of quadratic constraints is function specific yet there are certain heuristics that can be utilized discussed below.
	
	\begin{definition}
		Given $\alpha\leq \beta$, function $\psi:\mathbb{R}\to \mathbb{R}$ lies in sector $[\alpha,\beta]$ if
		\begin{equation*}
		(\psi(x)-\alpha x)(\beta x-\psi(x))\geq 0, ~~~\forall x \in \mathbb{R}.
		\end{equation*}
	\end{definition} 
	
\begin{figure}[t]
		\centering
		\includegraphics[scale=0.27]{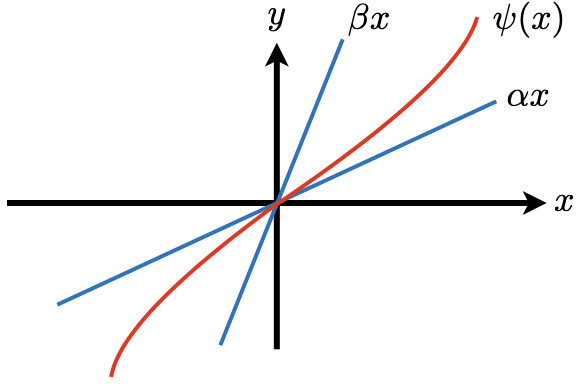}
		\caption{Illustration of sector bounds for nonlinearity with $\alpha\leq \frac{\psi(x)}{x}\leq \beta$.}
		\label{fig1}
	\end{figure}

	The interpretation of sector $[\alpha,\beta]$ is that $y=\psi(x)$ lies in the region formed by lines $y=\alpha x$ and $y=\beta x$ passing through origin (see Fig.~\ref{fig1}).
	
% 	 \begin{figure}[h!]
% 		\centering
% 		\includegraphics[scale=0.17]{SBT.png}
% 		\caption{Illustration of sector bounds for nonlinearity, i.e., $\alpha\leq \frac{\psi(x)}{x}\leq \beta$.}
% 		\label{fig1}
% 	\end{figure} 
	 
	Local sector constraints can also be defined for vector-valued functions $\phi : \mathbb{R}^{n_\phi} \rightarrow \mathbb{R}^{n_\phi}$. These local sectors can be concatenated in the form of vectors $\alpha_\phi,\beta_\phi \in \mathbb{R}^{n_{\phi}}$. 
	\begin{lemma}[\cite{yin2021imitation}]\label{lemma3.3}
		Let $\alpha_\phi$, $\beta_\phi$, $\underline{x}$, $\bar{x}$ $\in \mathbb{R}^{n_\phi}$ be given with $\alpha_\phi \leq \beta_\phi$. And $\phi$ satisfies the local sector $[\alpha_\phi,\beta_\phi]$ coordinate-wise  for all $x\in [\underline{x},\bar{x}]$. If $\mu\in \mathbb{R}^{n_\phi}$ with $\mu\geq 0$ then:
		\begin{gather}\label{QCac}
		\small{\begin{bmatrix} x  \\  \phi(x)  \end{bmatrix}^\top
		\begin{bmatrix}
		-2A_{\phi}B_\phi M & (A_{\phi}+B_{\phi})M \\ (A_{\phi}+B_{\phi})M & -2M
		\end{bmatrix}
		\begin{bmatrix} x  \\  \phi(x)  \end{bmatrix}}
		\geq 0
		\end{gather}
		where $A_{\phi}=diag(\alpha_\phi)$, $B_{\phi}=diag(\beta_\phi)$, and $M=diag(\mu)$.
	\end{lemma}
	
	\subsection{Admissibility analysis of NN}
	Based on QC abstractions and $\mathcal{S}$-procedure, the following result provides the admissibility condition of a fixed NN\cite{fazlyab2020safety}.
	\begin{thm}
		Consider a two-layer NN $\Psi:\mathbb{R}^{n_x} \to \mathbb{R}^{n_y}$ described by \eqref{nn1}, with nonlinear activation function sector bounded as in  \eqref{QCac}. Consider the matrix inequality
		\begin{equation}\label{lmi1}
		M_{\mathcal{X}}(P)+M_{\mathcal{Y}}(S)+M_{\Psi}(Q)\preceq 0
		\end{equation}
		where
		\begin{align*}
		M_{\mathcal{X}}(P) &=  \begin{bmatrix} I_{n_0} & 0 & 0 \\ 0 & 0  &  1 \end{bmatrix}^{\top}
		P
		\begin{bmatrix} I_{n_0} & 0 & 0 \\ 0 & 0  &  1 \end{bmatrix} \\
		M_{\Psi} (Q) &=  \textcolor{black}{\begin{bmatrix} \star \end{bmatrix}^\top}
		\scriptsize{\begin{bmatrix}
		-2A_{\phi}B_\phi M & (A_{\phi}+B_{\phi})M \\ (A_{\phi}+B_{\phi})M & -2M
		\end{bmatrix} \begin{bmatrix} W^{0} & 0 & b^{0} \\ 0 & I_{n_1} & 0 \\0 & 0 &  1 \end{bmatrix}} \\
		M_{\mathcal{Y}}(S) &=  \begin{bmatrix} I_{n_0} & 0 & 0 \\ 0 & W^{1^\top} & 0 \\ 0 & b^{1^\top} &  1 \end{bmatrix}
		S
		\begin{bmatrix} I_{n_0} & 0 & 0 \\ 0 & W^{1} & b^{1} \\ 0 & 0 &  1 \end{bmatrix}
		\end{align*}
		%with 
		%\begin{gather}
		%P(\Gamma) =  \begin{bmatrix}
		%H^T\Gamma H & -H^T\Gamma h\\ -h^T\Gamma H & h^T \Gamma h
		%\end{bmatrix}
		%\end{gather}
		%\begin{gather}
		%S =  \begin{bmatrix}
		%0 & 0& 0\\0 &-H_{y}^{T}H_y & H_{y}^{T}h_y\\0& h_{y}^{T}H_y& -h_{y}^{T}h_{y}
		%\end{bmatrix}
		%\end{gather}
		If \eqref{lmi1} is feasible for $(P,Q,S)$, then $\Psi(x)\in \mathcal{S}(x)$ for all $x\in \mathcal{X}$.
	\end{thm}
	
	 The above theorem implies that for a given NN, if \eqref{lmi1} holds, then we can certify the admissibility. However, it is seldom the case that an NN learned with an unconstrained approach satisfies the specified constraints. To reliably learn an admissible NN, it seems straightforward to impose the specifications as constraints. Nevertheless, the analysis condition \eqref{lmi1} is nonconvex with respect to both NN weights and multipliers $(P,Q,S)$, thus rendering the problem computationally intractable to solve.

	\section{Convexified learning under specifications}
	The key idea is to reparametrize the NN such that the condition becomes convex in a transformed space. To streamline the presentation, we derive our results for a two-layer NN with $l=1$ hidden layer, that is $\Psi(x)=W^1\phi(W^0x+b^0)+b^1$. In this case, equation \eqref{nn1} can be rewritten as
	\begin{gather}\label{1nn}
	\begin{bmatrix} v^0\\ \Psi(x)  \end{bmatrix}
	=
	N\begin{bmatrix} x \\ x^1  \end{bmatrix}+ \begin{bmatrix} b^0 \\ b^1  \end{bmatrix}\\
	x^1=\phi (v^0),
	\end{gather}
	where matrix $N$ depends on the weights as
	\begin{gather}\label{19}
	N=
	\begin{bmatrix}N_{vx} & N_{vx^1} \\ N_{\Psi x} & N_{\Psi x^{1}} \end{bmatrix}=
	\begin{bmatrix}W^0 & 0 \\ 0 & W^1 \end{bmatrix}.
	\end{gather}
	
	\subsection{Loop transformation}\label{looptrasnform}
	Loop transformation is a standard linear fractional transformation manipulation in  the control literature\textcolor{black}{\cite{yin2021imitation}}. Through loop transformation, we obtain a new representation that convexifies the learning condition without imposing restrictions on sector bounds $\alpha_\phi$ and $\beta_\phi$ of the activation function. In particular, loop transformation normalizes the nonlinearity $\tilde{\phi}$ to lie in the sector $[-1_{n_{\phi}\times 1},1_{n_{\phi}\times 1}]$. Thereby $\tilde{x}^{1}=\tilde{\phi}(v^{0})$ satisfies the quadratic constraint
	\begin{gather}\label{QCac2}
	\begin{bmatrix} v^{0}  \\  \tilde{x}^{1}  \end{bmatrix}^\top
	\begin{bmatrix}
	M & 0 \\0 & -M
	\end{bmatrix}
	\begin{bmatrix} v^{0}  \\  \tilde{x}^{1}  \end{bmatrix}
	\geq 0,~\forall v^{0}\in[\underline{v},\bar{v}]
	\end{gather}
    where $~M=diag(\mu)$. The input to $N$ is transformed by the algebraic equation
	\begin{equation}\label{T1}
	x^1=\frac{B_\phi- A_\phi}{2}\tilde{x}^1+\frac{B_\phi+ A_\phi}{2}v^0
	\end{equation}   
	Substituting \eqref{T1} in \eqref{1nn}, we get 
	\begin{equation}\label{T3}
	v^0= N_{vx}x+N_{vx^1}\left( \frac{B_\phi -A_\phi}{2}\tilde{x}^1 +\frac{A_\phi +B_\phi}{2}v^0\right) +b^0
	\end{equation}
	\begin{equation}\label{T4}
	\textcolor{black}{\Psi(x)}=N_{\Psi x}x+N_{\Psi x^{1}}\left( \frac{B_\phi -A_\phi}{2}\tilde{x}^1 +\frac{A_\phi +B_\phi}{2}v^0\right)+b^1.
	\end{equation}
	By solving \eqref{T3}, we obtain the expression for $v^0$, 
	\begin{equation}
	v^0=\left( I-C_2\right) ^{-1} N_{vx}x+\left( I-C_2\right) ^{-1}C_1\tilde{x}^1 +\left( I-C_2\right)^{-1} b^0
	\end{equation}
	Substituting $v^0$ in \eqref{T4} yields
	\begin{equation}
	\begin{split}
	\textcolor{black}{\Psi(x)} =&\left( N_{\Psi x}+C_4\left( I-C_2\right) ^{-1}N_{vx}\right) x+C_4\left( I-C_2\right) ^{-1}b^0\\&+\left( C_3+C_4\left( I-C_2\right) ^{-1}C_1\right) \tilde{x}^{1}+b^1,
	\end{split}
	\end{equation}
	with\\
	\begin{equation}\nonumber
	    C_1=N_{vx^1}\frac{ B_\phi -A_\phi}{2},~~ C_2=N_{vx^1}\frac{ B_\phi +A_\phi}{2},
	\end{equation}
	\begin{equation}\nonumber
	    C_3=N_{\Psi x^{1}}\frac{ B_\phi -A_\phi}{2},~~ C_4=N_{\Psi x^{1}}\frac{ B_\phi +A_\phi}{2}.
	\end{equation}\\
	After applying the loop transformation, the new representation of the NN is equivalent to 
	\begin{gather}\label{A1}
	\begin{bmatrix} v^0\\  \textcolor{black}{\Psi(x)} \end{bmatrix}
	=
	\tilde{N}\begin{bmatrix} x \\ \tilde{x}^1  \end{bmatrix}+ \begin{bmatrix} \tilde{b}^0 \\ \tilde{b}^1  \end{bmatrix}\\
	\tilde{x}^1=\tilde{\phi} (v^0).
	\end{gather}
	where 
	$\tilde{b}^0=\left( I-C_2\right) ^{-1} b^{0}$, $\tilde{b}^1=C_2\left( I-C_2\right) ^{-1}b^0+b^1$, and 
	\begin{align}\label{27}
	    \begin{split}
	          \tilde{N}&=\begin{bmatrix} \left( I-C_2\right) ^{-1}N_{vx} & \left( I-C_2\right) ^{-1}C_1 \\  N_{\Psi x}+C_4\left( I-C_2\right) ^{-1}N_{vx}& C_3+C_4\left( I-C_2\right) ^{-1}C_1 \end{bmatrix}\\&:=\begin{bmatrix}
	          \textcolor{black}{\tilde{N}_{vx}} &\textcolor{black}{\tilde{N}_{vx^{1}}}\\\textcolor{black}{\tilde{N}_{\Psi x}}&\textcolor{black}{\tilde{N}_{\Psi x^{1}}}
	          \end{bmatrix}
	    \end{split}
	\end{align}
	It can be seen that $\tilde{N}$ is in general a nonlinear function of $N$. To solve the equation, an ADMM algorithm is developed in \cite{yin2021imitation}. Also, it is important to note that $\tilde{N}$ depends  indirectly on $N$ through the sector bounds $\left( A_\phi,B_\phi \right) $. Specifically, suppose both $N$ and the state bounds are given. Then $\tilde{N}$ is constructed by: $(i)$ propagating \textcolor{black}{the bounds on $x$} through NN to compute bounds $\underline{v},\bar{v}$ on the activation inputs, $(ii)$ compute local sector bounds $A_{\phi},B_{\phi}$ consistent with the activation bounds, and $(iii)$ performing steps to compute $\tilde{N}$ from $(N, A_{\phi}, B_{\phi})$. Hereafter we treat $\tilde{N}$ as decision variable instead of $N$ (i.e., $\tilde N$ is the reparametrization of $N$).
	
	\subsection{Admissibility condition after loop transformation}
	Now we analyze the admissibilty of NN after loop transformation. Consider input and output sets to be ellipsoids. Based on the new representation of NN, the matrix inequality in \eqref{lmi1} can be rewritten as
	\begin{equation}\label{lmi21}
	\tilde{M}_\mathcal{X}+\tilde{M}_{\Psi}+\tilde{M}_{\mathcal{Y}}\preceq 0,
	\end{equation}
	with
	\begin{align}
	\tilde{M}_{\mathcal{X}} &=  \small{\begin{bmatrix} I_{n_0} & 0 \\ 0 & 0 \\ 0 &  1 \end{bmatrix}
	\begin{bmatrix}
	- \lambda A^\top  A & - \lambda A^\top b\\ -\lambda b^\top  A & \lambda(1-b^\top b)
	\end{bmatrix}
	\begin{bmatrix} I_{n_0} & 0 \\ 0 & 0 \\ 0 &  1 \end{bmatrix}^\top} \label{m11} \\
	\tilde{M}_{\Psi} &=  \textcolor{black}{\begin{bmatrix} \star \end{bmatrix}^\top}
	\begin{bmatrix}
	M & 0 & 0\\0 &-M &0\\0&0&0
	\end{bmatrix}
	\begin{bmatrix} \tilde{N}_{vx} & \tilde{N}_{vx^{1}} & \tilde{b}^{0} \\ 0 & I_{n_1} & 0 \\0 & 0 &  1 \end{bmatrix} \label{m21} \\
	\tilde{M}_{\mathcal{Y}} &=  \small{\textcolor{black}{\begin{bmatrix} \star \end{bmatrix}^\top}
	\begin{bmatrix}
	0 & 0& 0\\0 &\textcolor{black}{C^{\top}C} & \textcolor{black}{C^{\top}d}\\0& \textcolor{black}{d^{\top}C}& \textcolor{black}{d^{\top}d-1}
	\end{bmatrix}
	\begin{bmatrix} I_{n_0} & 0 & 0 \\ \tilde{N}_{\Psi x} & \tilde{N}_{\Psi x^{1}} & \tilde{b}^{1} \\ 0 & 0 &  1 \end{bmatrix}} \label{m31}
	\end{align}
	By substituting \eqref{m11}--\eqref{m31} in \eqref{lmi21}, and after simplification, we obtain:
	\begin{align}
	\begin{split}
	\bigg[
	\star
	\bigg]^{\top}&
	\begin{bmatrix}
	M &0\\0&I_{y}
	\end{bmatrix}
	\begin{bmatrix}
	\tilde{N}_{vx} &\tilde{N}_{vx^1} &\tilde{b}^{0}\\
	 \textcolor{black}{C\tilde{N}_{\Psi x}} & \textcolor{black}{C\tilde{N}_{\Psi x^{1}}}& \textcolor{black}{C\tilde{b}^{1}} +d
	\end{bmatrix}\\&-
	\begin{bmatrix}
	\lambda A^\top A& 0& 	\lambda A^\top b \\0& M&0\\\lambda b^\top A&0&\lambda(b^{\top}b-1)+1
	\end{bmatrix}
	\preceq 0.
	\end{split}
	\end{align}	
	Applying Schur complements yields an equivalent condition: 
	\begin{gather}\label{35}
	\small{\begin{bmatrix}
	\lambda A^\top A&0& \lambda A^\top b&\tilde{N}^\top_{vx}&\tilde{N}^{\top}_{\Psi x}C^{\top}\\
	0&M&0&\tilde{N}_{vx^1}^{\top}&\tilde{N}_{\Psi x^{1}}^{\top}C^{\top}\\
	\lambda b^{\top}A&0&\lambda(b^{\top}b-1)+1&\tilde{b}^{0\top}&\tilde{b}^{1\top}C^{\top}+d^{\top}\\\tilde{N}_{vx}&\tilde{N}_{vx^1}&\tilde{b}^{0}&M^{-1}&0\\C\tilde{N}_{\Psi x}&C\tilde{N}_{\Psi x^{1}} &C\tilde{b}^{1}+d &0&I_{n_y}
	\end{bmatrix}\succeq 0}
	\end{gather}
	
	 The inequality in \eqref{35} is linear in NN weights and bias vectors, but still nonconvex in \textcolor{black}{$M$}. Now multiply \eqref{35} from the left and right by $diag\left( \left[\begin{smallmatrix}
	I_{n_x}&0\\0&\textcolor{black}{M}^{-1}
	\end{smallmatrix}\right],I_{1+n_1+n_y}\right)$ to get the required convex condition
	\begin{gather}\label{convex}
	\small{\begin{bmatrix}
	\lambda A^\top A&0&\lambda A^\top b&\tilde{N}^\top_{vx}&\tilde{N}^{\top}_{\Psi x}C^{\top}\\0&Q_{1}&0&L_{1}^{\top}&L_{2}^{\top}C^{\top}\\\lambda b^{\top}A&0&\lambda(b^{\top}b-1)+1&\tilde{b}^{0}&\tilde{b}^{1\top}C^\top+d^{\top}\\\tilde{N}_{vx}&L1&\tilde{b}^{0}&Q_1&0_{n_1\times n_{y}}\\C\tilde{N}_{\Psi x}&CL_{2}&C\tilde{b^1}+d &0_{n_y\times n_1}&I_{n_y}
	\end{bmatrix}\succeq 0}
	\end{gather}
	where
	$Q_1=(M)^{-1}\succ0$, $L_{1}=\tilde{N}_{vx1}Q_1$, and $L_2=\tilde{N}_{\Psi x^{1}}Q_{1}$.
	Now the above constraint \eqref{convex} is convex in the decision variables $Q_1,\lambda, L_1,L2, \tilde{N}_{vx}, \tilde{N}_{\Psi x}, \tilde{b}^{0}$, and $\tilde{b}^{1}$. As a result, we can efficiently search over the admissible NN parameters by imposing this condition during learning.
	
\subsection{Algorithm}
The learning procedure involves finding a feasible solution to the LMI condition \eqref{convex}, and recovering the NN parameters from the numerical solutions. It evolves as follows:
\begin{itemize}
    \item [1.] Formulate and abstract the provided input-output specifications according to Sections III-A and III-B.
    \item [2.] Numerically solve for a feasible solution $(L_1, L_2, Q_1)$ to the LMI \eqref{convex}.
    \item [3.] If a feasible solution is found successfully, compute the NN weights by solving \eqref{27}.
\end{itemize}

If a feasible solution is found in step 2, then by Theorem \ref{thm:main}, the corresponding NN recovered in step 3 certifiably meets the input-output specifications. Nevertheless, infeasibility of \eqref{convex} in general does not imply the emptiness of the admissible set---a limitation due to the potential conservativeness of convex relaxation approaches.

\subsection{Multi-layer neural network} The extension to a multi-layer NN is straightforward. Define $x=[x^{0},x^{1},\dots,x^{l}]$, where $l\geq1$ is the number of hidden layers, and $x^{k}=E^{k}x$ for $k=0,\dots,l$, where $E^{k}$ is the entry selector matrix. Also, denote
	\begin{align*}
	A&=
	  \small{\begin{bmatrix}
	W^{0}&0&0&\dots&0&0\\
	0&W^{1}&0&\dots&0&0\\
	.&.&W^{2}&\dots&.&.\\
	.&.&.&\dots&.&.\\
	.&.&.&\dots&.&.\\0&0&0&\dots&W^{l-1}&0
	\end{bmatrix}}  \\
	B&=
	  \small{\begin{bmatrix}
	0&I_{n_{1}}&0&\dots&0&0\\
	.&.&I_{n_{2}}&\dots&.&.\\
	.&.&.&\dots&.&.\\
	.&.&.&\dots&.&.\\0&0&0&\dots&0&I_{n_l}
	\end{bmatrix}}, 
	    b=
	\small{\begin{bmatrix}
	b^{0}\\
	b^{1}\\.\\.\\.\\ b^{l-1}
	\end{bmatrix}}
	\end{align*}\\
	The following result provides a convex condition for learning a multi-layer NN under specifications.
\begin{thm}
\label{thm:main}
    Consider a multi-layer NN $\Psi:\mathbb{R}^{n_x} \to \mathbb{R}^{n_y}$ described by \eqref{nn1}, with nonlinear activation function sector bounded as in  \eqref{QCac}. Consider the matrix inequality
		\begin{equation}\label{mlmi}
		\tilde{M}_{\mathcal{X}}(P)+\tilde{M}_{\mathcal{Y}}(S)+\tilde{M}_{\Psi}(Q)\preceq 0
		\end{equation}
		where
		\begin{align*}
		\tilde{M}_{\mathcal{X}}(P) &=  \begin{bmatrix} E^{0} & 0 \\  0  &  1 \end{bmatrix}^{\top}
		P
	\begin{bmatrix} E^{0} & 0 \\  0  &  1 \end{bmatrix} \\
		\tilde{M}_{\Psi} (Q) &=  \begin{bmatrix} \tilde{A}& \tilde{b} \\   B & 0 \\ 0 &  1 \end{bmatrix}^{\top}
				Q
		\begin{bmatrix} \tilde{A}& \tilde{b} \\   B & 0 \\ 0 &  1 \end{bmatrix} \\
		\tilde{M}_{\mathcal{Y}}(S) &=  \begin{bmatrix} E^{0} & 0 \\  \tilde{W}^{l^T}E^{l} & \tilde{b}^{l} \\ 0 &   1 \end{bmatrix}^{\top}
		S
		\begin{bmatrix} E^{0} & 0 \\  \tilde{W}^{l^T}E^{l} & \tilde{b}^{l} \\ 0 &   1 \end{bmatrix}.
		\end{align*}
		 	If \eqref{mlmi} is feasible for $(P,Q,S)$, then $\Psi(x)\in \mathcal{S}(x)$ for all $x\in \mathcal{X}$. \textcolor{black}{Here $(\tilde{A},\tilde{b})$ are transformed weights matrix and bias vector, obtained by the procedure explained in Section \ref{looptrasnform}.}
\end{thm}

\section{Numerical Experiments}
To validate the proposed method, we considered a reachability problem. For a given input set, the specification dictates that the NN output should lie within a set. Our task is to learn a set of NN weights and bias vectors satisfying such requirement. We implemented our algorithm in MATLAB and solved the LMI condition with SDPT3~\cite{Toh}. For the first experiment, we considered two pairs of ellipsoids of the same size as the input-output sets to learn a NN with dimensions $n_{x}=2$ and $n_y=2$ with $n_1=10$ hidden neurons. The input and output sets are shown in Fig.~2 (note that our method works in a wide range of positions and we only showcase one of typical examples here). To test the admissibility of the learned NN, we randomly generated 500 points in the input sets and propagated them through the NN. As expected, all the output points lied within the desired output sets.    
	
	\begin{figure}[t]
		\centering
		\includegraphics[scale=0.3]{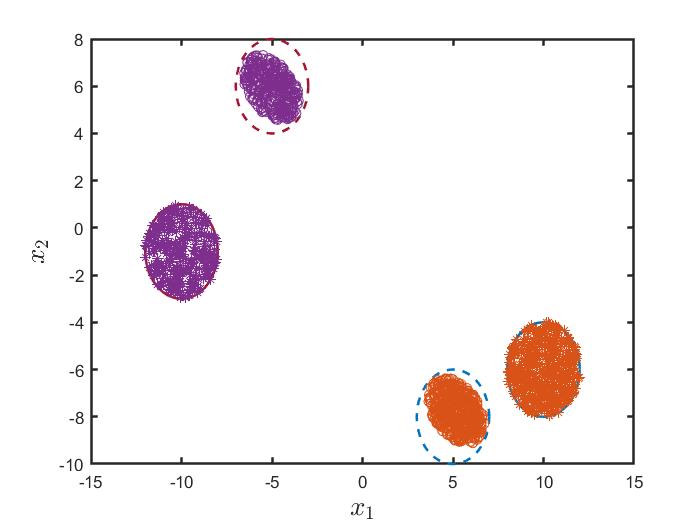}
		\vspace*{-1em}
		\caption{NN mapping with 10 hidden neurons.}
		\label{fig2}
	\end{figure} 
	
	\begin{figure}[t]
		\centering
		\includegraphics[scale=0.3]{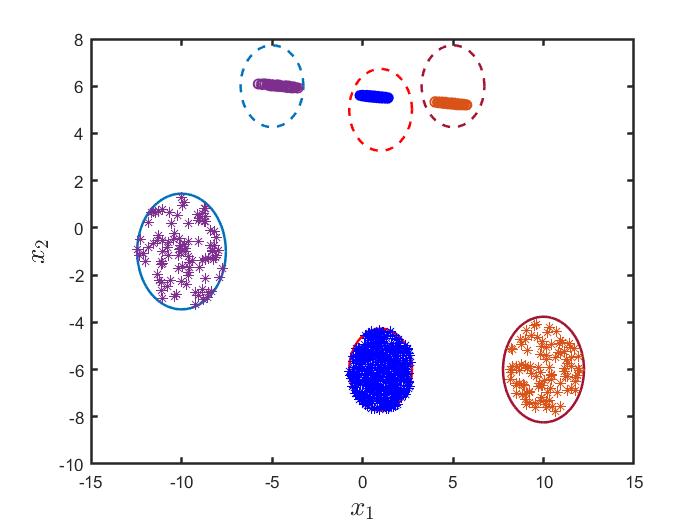}
		\vspace*{-1em}
		\caption{NN mapping with 5 neurons.}
		\label{fig3}
	\end{figure} 
	
	To ``stress test'' the proposed method, we conducted another experiment where we considered 3 pairs of input-output sets, but reduced the number of hidden neurons to 5. We also changed the sizes of input/output sets. A successful attempt is shown in Fig.~3. In general, as we increased the number of hidden layers, we observed a higher likelihood of finding a feasible solution, which is aligned with the universal approximation theorem \cite{D1}. Further quantification of the representation capacity informed by our admissibility condition is an interesting topic for future research.

	%%%%%%%%
	
	\section{Conclusion and Future Directions}
We address the challenge of learning NNs that certifiably satisfy input-output specifications. To tractably search for admissible weights, we derive a convex inner approximation to the nonconvex set of all admissible parameters. By abstracting the nonlinear specifications and activation functions with QCs, and applying the technique of loop transformation, we are able to derive a convex condition for a multi-layer NN that can be solved via SDP. The theoretical construction is verified by numerical experiments for a reachability-type problem. Building on the present work, there are several directions that we are currently pursuing, including (i) addressing more general forms of specifications, including those that can be approximated by QCs and those with internal dynamics; (ii) extending the theory to address convolutional neural networks, which have wide applications in extracting temporal and spatial correlations within data; and (iii) deriving learning-theoretic guarantees for a sample-based approach to solving problems with a large number of input-output specifications (that would otherwise been challenging to solve within a single SDP).

\balance

\end{document}